\documentstyle[epsfig,aps,prl,bbm,twocolumn,doublespace,12pt]{revtex}
\renewcommand{\theequation}{\arabic{equation}}
\begin{document}
%
%
%
%
\onecolumn
\title{The Topology of $SU(\infty)$ and the Group
of Area-Preserving Diffeomorphisms of a
Compact 2-manifold}
\author{John Swain}
\address{Department of Physics, Northeastern University, Boston, MA 02115, USA\\
email: john.swain@cern.ch}
\date{April 29, 2004}
\maketitle

\begin{abstract}
\section*{\bf Abstract}
Given the interest in relating the large $N$ limit of
$SU(N)$ to groups of area-preserving diffeomorphisms,
we consider the topologies of these groups and show that
both in terms of homology and homotopy, they are
extremely different. Similar conclusions are drawn for
other infinite dimensional classical groups.
\end{abstract}
 
\section{Introduction}

Groups of area-preserving diffeomorphisms and their Lie algebras
have recently been the focus of much attention in the physics 
literature.    
Hoppe \cite{Hoppe1} has shown that in a suitable basis, the Lie algebra of 
the group $SDiff(S^2)$ of area-preserving diffeomorphisms of a
sphere tends to that
of $SU(N)$ as $N\rightarrow\infty$.
This has obvious interest in 
connection with gauge theories of $SU(N)$ for large N. 
The idea of $SU(N)$ for finite $N$ as
an approximation to the group of area-preserving diffeomorphisms has also been
used in studies of supermembranes \cite{supermem1,supermem2,supermem3}, and
in particular has been used to argue for their instability.

The limiting procedure as $N\rightarrow\infty$ is delicate, and in particular, 
the need to take the limit in a particular basis makes one immediately 
wary as to how this result should be interpreted. In fact Hoppe and
Schaller \cite{Hoppe2}
have shown that there are infinitely many pairwise non-isomorphic Lie algebras,
each of which tends to $su(\infty)$, the Lie algebra of $SU(\infty)$,
as $N\rightarrow\infty$.
The authors of references \cite{supermem2},\cite{supermem3}, and \cite{mylimits}
have especially
emphasized the difficulties in relating such infinite limits with Lie algebras
of area-preserving diffeomorphisms.
Various authors
have considered special limits and/or large-N limits of
other classical Lie algebras \cite{Pope,Pope2,Wolski,deWit,Vassilevich}
as relevant for 2-manifolds other than spheres. In general, the approaches
in the literature concentrate on the Lie algebras of the groups of interest, 
though there is a global viewpoint showing the
inequivalence  $SDiff(S^2)$ and a certain action of $SU(N)$ provided
by the Majorana representation \cite{Majorana} of states of a spin-$j$ system by sets of
points on a sphere \cite{mypaper}.
 
The purpose of this Letter is to examine the topology of the groups of
area-preserving diffeomorphisms of compact 2-manifolds and make a comparison 
to the topology of $SU(N)$ both for finite $N$ and in the limit of 
$N\rightarrow\infty$. In addition, we examine the topology of other
infinite limits of classical groups and compare to $SDiff(M)$.

\section{Central Extensions and Topology}

The Lie algebra corresponding to the $SDiff(M)$ is the Lie algebra of divergence-free
vector fields. 
There have been discussions in the literature of the central extensions allowed to this 
algebra \cite{Pope}. The discussion is based on
the observation that divergence-free vector fields are, by the Poincar\'e lemma, locally
given by the curl of some other vector field. Globally, however, there are vector fields
whose divergence vanishes but which are not the curl of some other vector field. These can
provide central extensions to the Lie algebra and they are labelled by the 
independent harmonic 1-forms on $M$. There are $2g$ such independent forms on a 
2-manifold of genus $g$ corresponding to the $2g$ elements of
$H^1(M,{\mathbbm{R}})$.   
In \cite{Pope} an attempt is made to identify two distinct large $N$ limits
of the Lie algebra of $SU(N)$, $su(\infty)$ and $su_{+}(\infty)$ with the
$SDiff(M)$ for $M$ a 2-torus and a sphere respectively, based on the observation
that the former admits central extensions, while the latter does not.   
In this letter we concern ourselves not with the Lie algebras of the
groups under consideration, but rather with the global topology of the
groups themselves.

\section{Topology of $SU(\infty)$}\label{sec:Top}

The results
on de Rham cohomology and homotopy groups of $SU(\infty)$ 
presented in this section are well-known, but will be needed to
compare with corresponding results for groups of area-preserving diffeomorphisms.
As far as de Rham cohomology
is concerned, $SU(N)$ looks like a product of odd-dimensional spheres
\cite{encyc}, so that $SU(2) \cong S^3$,
$SU(3) \cong S^3 \ast S^5$, $SU(4) \cong S^3 \ast S^5\ast S^7$, and 
$SU(N) \cong S^3 \ast S^5\ast S^7\cdots\ast S^{2N-1}$.
Equivalently, the de Rham cohomology ring is $H^\ast(SU(N),{\mathbbm{R}}) = $
$\wedge_{\mathbbm{R}}(x_3,x_5,x_7,\ldots,x_{2N-1})$ where $deg(x_i) = i$.
The homotopy groups are
given by Bott's celebrated periodicity theorem :
$\pi_n(SU(\infty)) = Z$ for n odd and $n>1$, and $\pi_n(SU(\infty)) = 0$ for n even.

\section{Topology of $SDiff(M)$}

In what follows we will restrict out attention to the identity component of
the diffeomorphism groups considered, so that, for example, by $Diff(M)$
we will mean the identity component of the group of diffeomorphisms of $M$.
This is not a real restriction as we show with the following argument.
If $SDiff_0(M)$ is the identity component of $SDiff(M)$ then we know that

\begin{equation}
SDiff(M)/SDiff_0(M) = D
\end{equation}

\noindent where $D$ is a discrete group, but from this we cannot 
conclude that $SDiff(M)$ is a direct or semidirect product
\cite{Oraifeartaigh,Siebenthal}. However, we can now use the exact sequence
(see, for example reference \cite{BT}) :

\begin{equation}
\pi_j(D) \rightarrow \pi_{j-1}(SDiff_0(M)) \rightarrow \pi_{j-1}(SDiff(M))
\rightarrow \pi_{j-1}(D)
\end{equation}

\noindent and the fact that $D$ is discrete to conclude
that $\pi_i(D)= 0$ for $i>0$ and thus
that $\pi_i(SDiff(M)) = \pi_i(SDiff_0(M))$ for $i>0$.

We start by noting that for any manifold $M$ (not necessarily
of two dimensions), $SDiff(M)$ is a deformation retract
of the group $Diff(M)$ of diffeomorphisms of a manifold \cite{Omori,Ebin}.
Thus $SDiff(M)$ and $Diff(M)$ have the same homotopy and same de Rham cohomology
(see, for example \cite{BT}). 

Now we need the following result 
\cite{Earl,Gramain}
from the mathematics literature on 
$Diff(M)$ for 2-manifolds :

{\it Theorem : For $M$ a compact 2-dimensional manifold, $Diff(M)$ is homotopy equivalent
to 

\begin{itemize}
\item[1.] $SO(3)$ for $M = S^2$ or the projective plane
\item[2.] $S^1\times S^1$ for $M = T^2 = S^1\times S^1$
\item[3.] $SO(2)$ for $M$ a Klein bottle (KB) , M\"obius band, disk, or $S^1\times I$ (I an interval)
\item[4.] a point otherwise
\end{itemize}
}

Combining this with the fact that $SDiff(M)$ and $Diff(M)$ have the same de Rham cohomology
and homotopy groups, we have the following :

{\it Theorem : For $M$ a compact 2-dimensional manifold, $SDiff(M)$ has the same de Rham cohomology
and homotopy groups as

\begin{itemize}
\item[1.] $SO(3)$ for $M = S^2$ or the projective plane
\item[2.] $S^1\times S^1$ for $M = T^2 = S^1\times S^1$
\item[3.] $SO(2)$ for $M$ a Klein bottle (KB), M\"obius band, disk, or $S^1\times I$ (I an interval)
\item[4.] a point otherwise
\end{itemize}
}

We will find it convenient to group the compact 2-manifolds into 4 classes corresponding to the
4 numbered items in the above list.

\section{Comparison of $SU(\infty)$ and $SDiff(M)$} 

It is straightforward now to find the de Rham cohomology of $SDiff(M)$ and compare it
to that of $SU(\infty)$. We list the results for different $M$ in table 1, where we give only one
representative from each of the classes above with the understanding, for example,
that the results for
$S^2$ apply also to the projective plane, etc. The table continues with vanishing
cohomology in higher dimensions for $SDiff(M)$ while $SU(\infty)$ continues to
have nonvanishing cohomology in higher dimensions as discussed in section \ref{sec:Top}.

We also present the homotopy groups for $SDiff(M)$ and $SU(\infty)$ in Table 2 with the same
understanding that we list only one
representative from each of the classes of $M$. The first column uses the 
homotopy groups of $SO(3)$ tabuled in \cite{encyc}. Clearly the columns
of Table 1 are all different, as are the columns of Table 2. 
Thus we find that $SDiff(M)$ is clearly not the same as $SU(\infty)$.

\section{Other infinite dimensional classical groups}

Claims have been made in the literature \cite{Pope,Pope2,Wolski,deWit,Vassilevich} that various other infinite classical 
groups can be approximations to $SDiff(M)$. From the foregoing discussion, it should be
clear that the same problem of disagreement in homotopy and cohomology will arise. We consider
several cases in the following  subsections. An excellent general reference
for the homotopy and cohomology of the classical groups is
reference \cite{encyc}.

\subsection{Large Orthogonal Groups}

The group $SO(2N)$ has the de Rham cohomology
ring 

\begin{eqnarray}
H^\ast(SO(2N),{\mathbbm{R}}) =  
\wedge_{\mathbbm{R}}(x_3,x_5,x_7,\ldots,x_{4N-5})
\end{eqnarray}

while

\begin{eqnarray}
H^\ast(SO(2N-1),{\mathbbm{R}}) =
\wedge_{\mathbbm{R}}(x_3,x_5,x_7,\ldots,x_{4N-5},x_{2N-1}) 
\end{eqnarray}

Thus $SO(\infty)$ has nonvanishing
cohomology in many dimensions greater than 2. Its homotopy groups are
$\pi_0(SO(\infty)) = Z_2$,\ $\pi_1(SO(\infty)) = Z_2$,\ 
$\pi_2(SO(\infty)) = 0$,\ 
$\pi_3(SO(\infty)) = Z$,\ 
$\pi_4(SO(\infty)) = 0$,\ 
$\pi_5(SO(\infty)) = 0$,\
$\pi_7(SO(\infty)) = Z$,\ 
and $\pi_{n+8}(SO(\infty)) = \pi_{n}(SO(\infty))$ for
$n>7$.

\subsection{Large Symplectic Groups}

The group $Sp(N)$ has the cohomology ring of 

\begin{eqnarray}
H^\ast(Sp(\infty),{\mathbbm{R}}) =
\wedge_{\mathbbm{R}}(x_3,x_5,x_7,\ldots,x_{4N-1}) 
\end{eqnarray}

and thus also has nonvanishing
cohomology in many dimensions greater than 2. Its homotopy groups are
$\pi_n(Sp(\infty)) = \pi_{n+4}(SO(\infty))$.

In both the cases of $SO(\infty)$ and $Sp(\infty)$ we see that there are nonvanishing
cohomology groups in dimensions over 2, and clear mismatches with
homotopy groups of $SDiff(M)$, whatever the choice of $M$. 

\subsection{Other Large Groups}

We recall \cite{BT} that $\pi_n(A\times B) = \pi_n(A) \times \pi_n(B)$, and the
Kunneth formula which states that
$H^n(A\times B,{\mathbbm{R}}) = \sum_{l+m=n} H^l(A,{\mathbbm{R}}) \otimes
H^m(B,{\mathbbm{R}})$. Thus, in general,
the same problems encountered for $SO(\infty)$, $SU(\infty)$, and $Sp(\infty)$ will
be met by any products of such groups. Similarly, products of 
finite dimensional groups with the infinite classical groups will not
be diffeomorphic to $SDiff(M)$ for any 2-dimensional manifold $M$.


\section{Conclusion}

It is now clear that in terms of topology, $SU(N)$ approximates the topology of $SDiff(M)$ very
poorly except in the rather trivial case of $M = S^2$ and $N=3$. In that case
the de Rham cohomology and homotopy are the same ($SU(2)$ is the two-fold cover of $SO(3)$). 
Limits of the other classical groups also differ in their topology in a 
similarly dramatic manner.
This suggests that great caution be applied in the approximation of 
area-preserving diffeomorphisms (or, in fact, any diffeomorphisms at all)
by limits of $SU(N)$ or other large classical groups.
It has been suggested that $SU(\infty)$ might describe a theory of membranes
of changing topology \cite{supermem3}. One might even speculate that
$SU(\infty)$ describes a theory of not only 2-dimensional objects, but one with
strings and point particles as well. 
In this case, the topological structure of $SU(\infty)$ might be
expected to be of importance in mediating transitions between objects of
different topologies (or even dimensions?) much as $\pi_3(SU(3)) = Z$ gives
the instantons which tunnel between different vacua in QCD.

\section{Acknowledgement}

The author would like to
my colleagues at Northeastern University and the National Science
Foundation for their support.

\clearpage
\vbox{
\begin{center}
\begin{tabular}{|c|c|c|c|c|c|}
\hline
 $k$ &
 $H^k(SDiff(S^2))$ &
 $H^k(SDiff(T^2))$  &
 $H^k(SDiff(KB))$ &
 $H^k(SDiff(M^g))$ &
 $H^k(SU(\infty))$ \\
\hline & & & & &\\[-10mm]
$1$ & $0$ & $\mathbbm{R}$ $\oplus$ $\mathbbm{R}$ & $\mathbbm{R}$ & $\mathbbm{R}^g$ & $0$ \\
\hline & & & & &\\[-10mm]
2 & 0 & 0 & 0 & 0 & 0 \\
\hline & & & & &\\[-10mm]
3 & ${\mathbbm{R}}$ & 0 & 0 & 0 & ${\mathbbm{R}}$ \\
\hline & & & & &\\[-10mm]
4 & 0 & 0 & 0 & 0 & 0 \\
\hline & & & & &\\[-10mm]
5 & 0 & 0 & 0 & 0 & ${\mathbbm{R}}$ \\
\hline
\end{tabular}
\end{center}

\centerline{Table 1. De Rham cohomology groups of $SDiff(M)$ and $SU(\infty)$.}
}

\clearpage
\small

\vbox{
\begin{center}
\begin{tabular}{|c|c|c|c|c|c|}
\hline & & & & &\\[-10mm]
 k &
 $\pi_k(SDiff(S^2))$ &
 $\pi_k(SDiff(T^2))$  &
 $\pi_k(SDiff(KB))$ &
 $\pi_k(SDiff(M^g))$ &
 $\pi_k(SU(\infty))$ \\
\hline & & & & &\\[-10mm]
1 & 0 & $\mathbbm{Z}$ $\oplus$ $\mathbbm{Z}$ & $\mathbbm{Z}$ & $\mathbbm{Z}$ & 0 \\
\hline & & & & &\\[-10mm]
2 & 0 & 0 & 0 & 0 & 0 \\
\hline & & & & &\\[-10mm]
3 & $\mathbbm{Z}$ & 0 & 0 & 0 & $\mathbbm{Z}$ \\
\hline & & & & &\\[-10mm]
4 & $\mathbbm{Z}_2$ & 0 & 0 & 0 & 0 \\
\hline & & & & &\\[-10mm]
5 & $\mathbbm{Z}_2$ & 0 & 0 & 0 & $\mathbbm{Z}$ \\
\hline & & & & &\\[-10mm]
6 & $\mathbbm{Z}_{12}$ & 0 & 0 & 0 & 0 \\
\hline & & & & &\\[-10mm]
7 & $\mathbbm{Z}_{2}$ & 0 & 0 & 0 & $\mathbbm{Z}$ \\
\hline & & & & &\\[-10mm]
8 & $\mathbbm{Z}_{2}$ & 0 & 0 & 0 & 0 \\
\hline & & & & &\\[-10mm]
9 & $\mathbbm{Z}_{3}$ & 0 & 0 & 0 & $\mathbbm{Z}$ \\
\hline & & & & &\\[-10mm]
10 & $\mathbbm{Z}_{15}$ & 0 & 0 & 0 & 0 \\
\hline & & & & &\\[-10mm]
11 & $\mathbbm{Z}_{2}$ & 0 & 0 & 0 & $\mathbbm{Z}$ \\
\hline & & & & &\\[-10mm]
12 & $\mathbbm{Z}_{2}\oplus \mathbbm{Z}_{2}$ & 0 & 0 & 0 & 0 \\
\hline & & & & &\\[-10mm]
13 & $\mathbbm{Z}_{12}\oplus \mathbbm{Z}_{3}$ & 0 & 0 & 0 & $\mathbbm{Z}$ \\
\hline & & & & &\\[-10mm]
14 & $\mathbbm{Z}_{84}\oplus \mathbbm{Z}_{2}\oplus \mathbbm{Z}_{2}$ & 0 & 0 & 0 & 0 \\
\hline & & & & &\\[-10mm]
15 & $\mathbbm{Z}_{2}\oplus \mathbbm{Z}_{2}$ & 0 & 0 & 0 & $\mathbbm{Z}$ \\
\hline
\end{tabular}
\end{center}

\centerline{Table 2. Homotopy groups of $SDiff(M)$ and $SU(\infty)$.}
}



\clearpage

\begin{thebibliography}{99}
\bibitem{Hoppe1} J. Hoppe, Proceedings, Constraints Theory
and Relativistic Dynamics (Florence, 1986) eds. G. Longhi and L Lusana (World
Scientific, Singapore, 1987) p. 267; Phys. Lett. {\bf B215} (1988) 706.
\bibitem{supermem1} B. deWit, J. Hoppe, and H. Nicolai,
Nucl. Phys. {\bf B305} (1988) 545.
\bibitem{supermem2} B. deWit, M. L\"uscher, and H. Nicolai,
Nucl. Phys. {\bf B320} (1989) 133.
\bibitem{supermem3} B. de Wit and H. Nicolai, Proceedings, 3$^{rd}$
Hellenic School on
Elementary Particle Physics (Corfu, 1989) eds. E. N. Argyres et al.
(World Scientific, Singapore) p777.
\bibitem{Hoppe2} J. Hoppe and P. Schaller, Phys. Lett. {\bf B237} (1990) 407.
\bibitem{mylimits} J. Swain, ``On the limiting procedure by which $SDiff(T^2)$ and
$SU(\infty)$ are associated'',  {\tt http://arXiv.org/abs/hep-th/0405002}.
\bibitem{Pope} C. N. Pope and K. S. Stelle, Phys. Lett. {\bf B226} (1989) 257.
\bibitem{Pope2} C. N. Pope and L. J. Romans, Class. Quantum. Grav. {\bf 7}
(1990) 97. 
\bibitem{Wolski} A. Wolski and J. S. Dowker,
J. Math. Phys. {\bf 32} (1991) 2304.
\bibitem{deWit} B. deWit, U. Marquard, and H. Nicolai,
Comm. Math. Phys. {\bf 128}
(1990) 39.
\bibitem{Vassilevich} D. V. Vassilevich,
Class. Quantum. Grav. {\bf 8} (1991) 2163.
\bibitem{Majorana} E. Majorana, Nuov. Cim. {\bf 9} (1932) 43.
\bibitem{mypaper} J. Swain, ``The Majorana representation of spins and
the relation between $SU(\infty)$ and $SDiff(S^2)$'', 
{\tt http://arXiv.org/abs/hep-th/0405004}.
\bibitem{Oraifeartaigh} L. O'Raifeartaigh, ``Group Structure of 
Gauge Theories'', Cambridge Press 1986, Cambridge.
\bibitem{Siebenthal} J. de Siebenthal, Comment. Math Helv. {\bf 31} (1956) 41.
\bibitem{Omori} H. Omori, ``On the group of diffeomorphisms on a compact manifold'', in 
Global Analysis, Proc. Symp. Pure Math. {\bf 15}, ed. S. S. Chern and S. Smale (Amer.
Math. Soc., Providence, RI, 1970) 167.
\bibitem{Ebin} D. Ebin and J. Marsden, Ann. Math. {\bf 92} (1970) 101.
\bibitem{BT} R. Bott and L. Tu, ``Differential Forms in Algebraic Topology'',
 Springer-Verlag (1982), New York.
\bibitem{Earl} C. J. Earl and J. Eells, Bull. Am. Math. Soc. {\bf 73} (1967) 557, and
               C. J. Earl and J. Eells, J. Diff. Geom. {\bf 3} (1969) 19.
\bibitem{Gramain} A. Gramain, ``Groupes des diff\'eomorphismes et espace de Teichm\"uller d'une surface'',
S\'eminaire Bourbaki 1972-1973, Springer Lecture Notes in Mathematics 383 (1974) 157, and
A. Gramain ``Le type d'homotopie du groupe des diff\'eomorphismes d'une surface compacte'', Ann. Sci. 
de l'Ecole Normale Sup\'erieure {\bf 6} (1973) 53.
\bibitem{encyc} S. Iyanaga and Y. Kawada(eds.), ``Encyclopedic Dictionary of
Mathematics'', The MIT Press (1977) (Appendix A).
\end{thebibliography}
\end{document}